\documentclass[a4,12pt]{article}
\usepackage{a4,epsfig}
\newcommand\eps{\varepsilon}

\newcommand\pt{\partial}
\newcommand\tx{\tilde x}
\newcommand\ty{\tilde y}
\newcommand\bcA{\bar{\cal A}}
\newcommand\balpha{\bar\alpha}
\newcommand\bbeta{\bar\beta}
\newcommand\bgamma{\bar\gamma}
\begin{document}
%\linenumbers
\title{Luminosity Loss due to Kicks and Mismatch\\
from radio-frequency breakdown in a Linear Collider}
\author{V. Ziemann\\
Department of Physics and Astronomy\\
Uppsala University, Uppsala, Sweden}
\date{June 21, 2017}
\maketitle
%%%%%%%%%%%%%%%%%%%%%%%%%%%%%%%%%%%%%%%%%%%%%%%%%%%%%%%%%%%%%%%%%%%%%%%
\begin{abstract}\noindent
%\abstract{
We calculate the geometric luminosity loss caused by filamentation of
transverse kicks, upright and skew quadrupolar errors due to discharges, 
so-called RF-breakdown, in the acceleration structures of a Linear 
Collider.
\end{abstract}
%\footnotetext{Corresponding author: V. Ziemann (volker.ziemann@physics.uu.se),
%Department Physics and Astronomy, Uppsala University, Box 516, 
%75120 Uppsala, Sweden.}
%
%%%%%%%%%%%%%%%%%%%%%%%%%%%%%%%%%%%%%%%%%%%%%%%%%%%%%%%%%%%%%%%%%%%%%%%
%
\section{Introduction}
The required high accelerating gradient of 100\,MV/m in the CLIC~\cite{CLIC}
linear accelerator causes surface electric fields which are in excess of 200\,MV/m 
and this occasionally causes discharges, so-called RF breakdown, in which
a plasma is generated leading to the ejection of electrons and ions from
accelerating structures~\cite{MAGNUS}. Moreover, the plasma effectively
generates an electric short circuit in the structure that causes the 
RF fields to be reflected~\cite{RF}. Recently, we analyzed the effect of
these discharges on the accelerated beam~\cite{ANDREA} and found that the
discharges cause a transverse kick of the beam as well as changing the 
beam size and therefore causing a betatron mismatch.
\par
In case that a discharge occurs early in the linear accelerator at lower 
energies we expect that the finite momentum spread of the beam will cause 
full filamentation~\cite{FILA} of the displacement or betatron mismatch, 
because particles kicked or otherwise displaced to different amplitudes
from those of a matched and centered beam will start betatron oscillations
at their amplitudes and eventually will be spread out over a matched 
phase space ellipse, which will cause the emittance to grow and the
distribution of particles in phase space to change. This effect is 
normally considered as phase space dilution due to injection mismatch 
in storage rings, but the same concepts can be applied to evaluate the 
effect of mismatch an transverse kicks on the luminosity in linear colliders.
\par
In order to calculate the reduction of the luminosity we need to take the
detailed shape of the distribution after filamentation into account, 
considering rms beam sizes only is insufficient, because the final
distributions are sometimes significantly different from Gaussian~\cite{TOR}.
\par
In the remainder of this report we first introduce the relevant beam 
dynamics concepts that are needed for to calculate these non-Gaussian
distributions. We start by investigating the the luminosity loss due to a 
transverse kick in one phase space dimension and then progress to the 
consequence of betatron mismatch and finally consider the result of a
localized skew quadrupolar error and finally summarize our findings in 
the conclusions.
%
%%%%%%%%%%%%%%%%%%%%%%%%%%%%%%%%%%%%%%%%%%%%%%%%%%%%%%%%%%%%%%%%%%%%%%%
%
\section{Normalized Phase Space}
In order to simplify the algebra, we start by introducing normalized phase
space $(\tx,\tx')$, denoted by  a tilde. It is related to normal transverse
phase space coordinates $(x,x')$ in an accelerator by the transformation
\begin{equation}\label{eq:nps}
\left(\begin{array}{c} x \\ x'\end{array}\right)
=
\left(\begin{array}{cc}
\sqrt{\beta} & 0 \\
-\alpha/\sqrt{\beta} & 1/\sqrt{\beta}
\end{array}\right)
\left(\begin{array}{c}\tx \\ \tx'\end{array}\right)
= {\cal A} 
\left(\begin{array}{c}\tx \\ \tx'\end{array}\right)
\end{equation} 
where $\beta$ and $\alpha$ are the normal Twiss parameters at a given location
in the magnet lattice. Here we also introduced the abbreviation $\cal A$ for 
the matrix appearing in the previous equation.
\par
In all our analysis we assume that the initial beam distribution is Gaussian
in all phase space coordinates. In particular, the matrix of central second 
moments $\bar\sigma$ , the so-called sigma matrix, can be expressed in terms of 
the matrix $\cal A$ and the emittance $\eps$ as
\begin{equation}\label{eq:sigma}
\bar\sigma = \eps {\bcA}{\bcA}^t
=\eps
\left(\begin{array}{cc}
\bbeta & -\balpha \\ -\balpha & \bgamma
\end{array}\right)
\end{equation}
where ${\bcA}^t$ denotes the transpose of $\bcA$ and with the condition 
$\bgamma= (1+\balpha^2)/\bbeta.$ Here the quantities with a bar denote those 
of the incoming beam. If the incoming beam is matched, we have $\bbeta=\beta$ 
and $\balpha=\alpha.$
\par
Propagating beam particles through a beam line consisting of linear elements
is accomplished by transfer matrices $R$ which can also be represented 
using the matrices $\cal A$ and normal $2\times 2$ rotation matrices
\begin{equation}\label{eq:Rnps}
R={\cal A}_f {\cal O}_{\mu} {\cal A}_i^{-1}
\qquad\mathrm{with}\quad
{\cal O}_{\mu} = 
\left(\begin{array}{rc}
\cos\mu & \sin\mu \\ -\sin\mu & \cos\mu
\end{array}\right)
\end{equation}
and ${\cal A}_i$ contains the Twiss parameters at the initial location of 
the beam line section represented by $R$ and ${\cal A}_f$ those of the final
location. The betatron phase advance is denoted by $\mu.$
\par
As a trivial example to illustrate the usefulness of the formalism we 
consider a periodic  system with ${\cal A}_f = {\cal A}_i$ and a matched 
beam where the initial beam matrix $\sigma_i$ is constructed from the 
same ${\cal A}_i$ according to eq.~\ref{eq:sigma}. In that case we 
calculate the final beam matrix $\sigma_f$ from
\begin{equation}
\sigma_f = R\sigma_iR^t 
= 
{\cal A}_i {\cal O} {\cal A}_i^{-1}
\eps {\cal A}_i {\cal A}_i^t 
\left({\cal A}_i {\cal O} {\cal A}_i^{-1}\right)^t
=\eps {\cal A}_i {\cal A}_i^t = \sigma_i
\end{equation}
and find that it is equal to the initial beam matrix $\sigma_i,$ as expected.
\par
The sigma matrix is a convenient tool to propagate the beam through a linear
array of beam line elements, but the beam is always assumed to be Gaussian.
In order to be able to handle more general distributions we start by considering
an initial multi-variate Gaussian particle distribution $\psi(x,x')$ in the 
particle coordinates $(x_1,x_2)=(x,x')$ which is given by
\begin{equation}\label{eq:psi}
\psi(x,x') = \frac{1}{2\pi\sqrt{\det\sigma}}\exp\left[-\frac{1}{2}
  \sum_{i=1}^2\sum_{j=1}^2 \sigma_{ij}^{-1}x_i x_j\right]\ .
\end{equation}
This distribution is normalized to unity and characterized by the sigma matrix 
$\sigma.$ This expression can be trivially generalized to more phase space 
coordinates. Note that we can write the inverse of a sigma matrix, using
eq.~\ref{eq:sigma} as
\begin{equation}
\sigma^{-1} = \frac{1}{\eps} \left({\cal A}^t\right)^{-1}{\cal A}^{-1}
= 
\frac{1}{\eps}
\left(\begin{array}{cc}
\gamma & \alpha \\ \alpha & \beta
\end{array}\right)
\end{equation}
which will prove convenient later on. 
\par
Any distribution function $\Psi_X(\vec X)$
depending on variables $\vec X$ can be transformed to new variables $\vec Y = f(\vec X)$ 
in the following way
\begin{equation}
\Psi_Y(\vec Y) = \frac{1}{\vert J_f(\vec Y,\vec X)\vert} \Psi_X\left(f^{-1}(\vec Y)\right)
\end{equation} 
where $\Psi_Y$ depends on the new variables $\vec Y.$ The inverse of the variable 
transformation $f$ is denoted by $f^{-1}$ and $J_f(\vec Y,\vec X)$ refers to the 
Jacobian of $f$. In the particular case where the variable transformation $f$ 
stems from a linear transformation described my matrix $R$ the distribution
function $\psi(x,x')$ in eq.~\ref{eq:psi} is transformed to new variables 
$(y,y')^t=R\,(x,x')^t$ by simply replacing the sigma matrix in eq.~\ref{eq:psi}
by the one transformed to the new coordinates, given by the familiar
expression $R\sigma R^t$.
\par
In the case where we change the variables of the distribution function to those
of normalized phase space using eq.~\ref{eq:nps} we find
\begin{equation}\label{eq:npsdist}
\psi(\tx,\tx') = \frac{1}{2\pi\eps}\exp\left[-\frac{\tx^2+\tx^{\prime 2}}{2\eps}\right]
\end{equation}
which describes a rotationally symmetric Gaussian in the two phase space variables
$\tx$ and $\tx'.$ Introducing action and angle variables $(J,\phi)$ in normalized 
phase space by 
\begin{equation}\label{eq:aa}
\tx = \sqrt{2J}\cos\phi\qquad\mathrm{and}\qquad \tx'=\sqrt{2J}\sin\phi 
\end{equation}
we can write the previous equation in the form
\begin{equation}\label{eq:exp}
\psi(J,\phi) = \frac{1}{2\pi\eps}\, e^{-J/\eps}
\end{equation}
where we have used the fact that the Jacobian of the transformation from $\tx$ and 
$\tx'$ to $J$ and $\phi$ has unit Jacobian and the expression for the matched 
beam in action-angle variables in normalized phase space is independent of the
angle variable. Note that propagation with a betatron phase advance $\mu$ 
increases the phase variable $\phi$ by $\mu$ and leaves the action variable
unaffected, the distribution in eq.~\ref{eq:exp} remains the same.
\par
We start by considering the effect of transverse kicks on the luminosity in the
following section.
%
%%%%%%%%%%%%%%%%%%%%%%%%%%%%%%%%%%%%%%%%%%%%%%%%%%%%%%%%%%%%%%%%%%%%%%%
%
\section{Transverse Kick}
In this section we address the effect of a localized transverse kick that
completely filaments on its journey down the linear accelerator. We assume 
that the initial beam is matched to the beam line, but all particles of 
the distribution receive a transverse kick of magnitude $\theta$ which 
changes the particle coordinates from $(x,x')$ to $(x, x'-\theta).$
Translated to normalized phase space this kick leads to the variable
transformation from $(\tx,\tx')$ to $((\tx,\tx'-\sqrt{\beta}\theta).$
Applying this coordinate transformation to the matched beam given
by eq.~\ref{eq:npsdist} we obtain the distribution for the kicked beam
\begin{equation}
\psi_k(\tx,\tx') = \frac{1}{2\pi\eps} 
   \exp\left[-\frac{\tx^2 + (\tx'-\sqrt{\beta}\theta)^2}{2\eps}\right] 
\end{equation}
and introducing the action angle variables given in eq.~\ref{eq:aa}
arrive at 
\begin{equation}\label{eq:kicked}
\psi_k(J,\phi) = \frac{1}{2\pi\eps} 
\exp\left[-\frac{J}{\eps} - \frac{\beta\theta^2}{2\eps}\right]
\exp\left[\frac{\theta\sqrt{2J\beta}\sin\phi}{\eps}\right]
\end{equation}
which reduces to the un-kicked distribution in eq.~\ref{eq:exp} in the 
case that $\theta=0.$
\par
Equation~\ref{eq:kicked} describes the distribution immediately after 
the kick. During its further passage down the linear accelerator we assume
that the beam filaments completely, which means that the particles located
at a given angle $\phi$ are evenly distributed over all phase angles,
which amounts to averaging the kicked distribution $\psi_k(J,\phi)$ in
eq.~\ref{eq:kicked} over all angles $\phi$ to arrive at the distribution 
$\psi_f(J)$ after filamentation, given by
\begin{equation}\label{eq:psif}
\psi_f(J) = \frac{1}{2\pi} \int_0^{2\pi} \psi_k(J,\phi) d\phi
= 
\frac{1}{2\pi\eps} \exp\left[-\frac{J}{\eps} - \frac{\beta\theta^2}{2\eps}\right]
I_0\left(\frac{\sqrt{2J\beta}\theta}{\eps}\right)   
\end{equation}
where we have used the integral representation of the modified Bessel function
$I_0(z)$ given by equation 9.6.16 in Ref.~\cite{ABRASTE} to evaluate the integral
over $\phi.$ This expression is similar to the one found for example in 
ref.~\cite{FISCHER}.
\par
%..............................................
\begin{figure}[tb]
\begin{center}
\epsfig{file=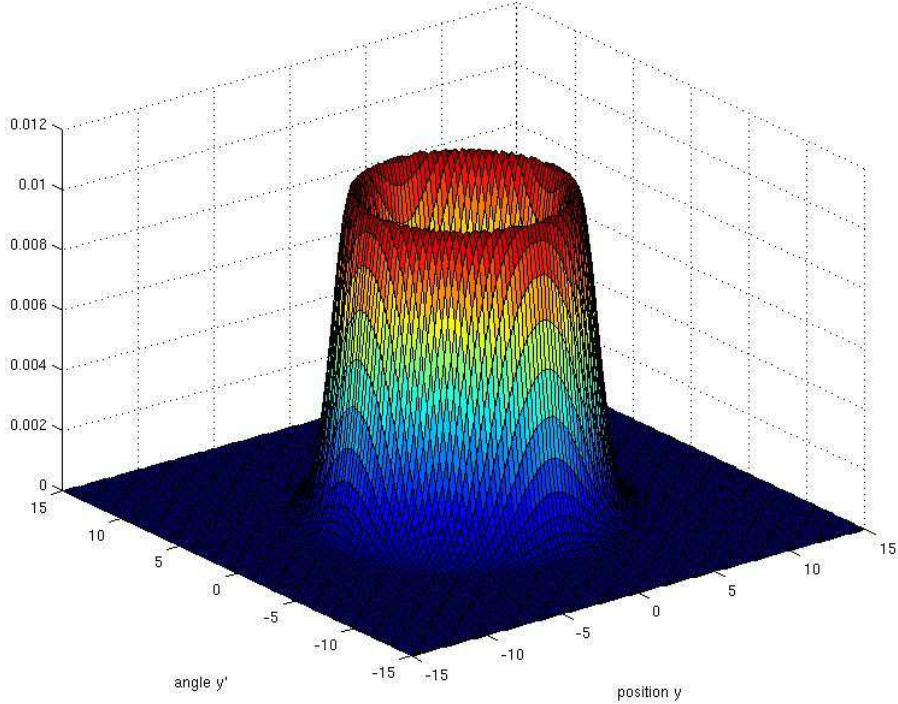,width=0.95\textwidth}
\end{center}
\caption{\label{fig:dist}Kicked distribution in normalized phase space after
filamentation. The kick amplitude corresponds to 6 times the angular divergence.}
\end{figure}
%..............................................
%..............................................
\begin{figure}[tb]
\begin{center}
\epsfig{file=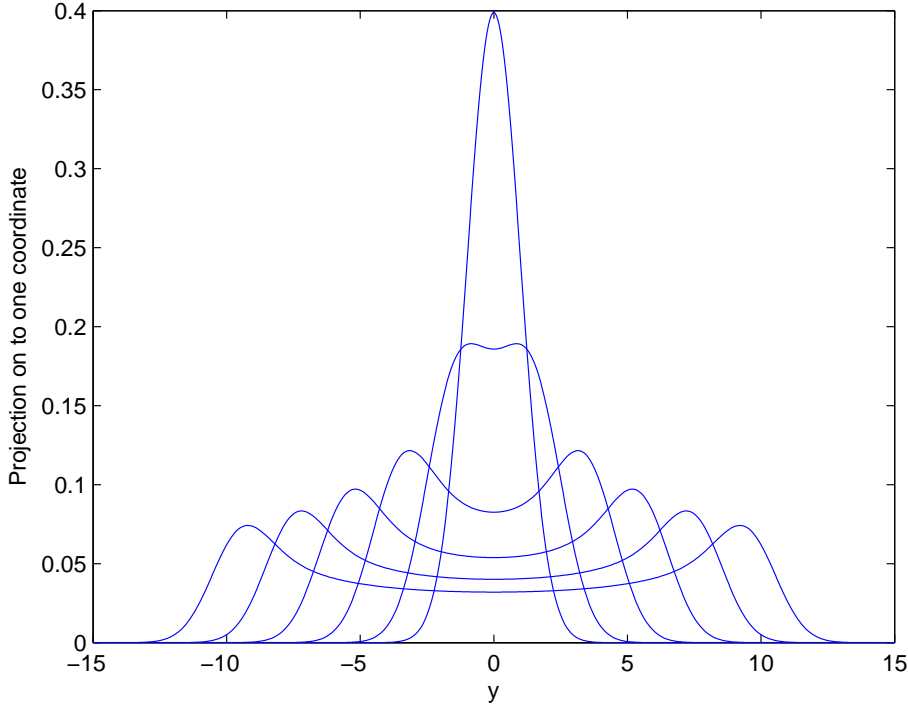,width=0.95\textwidth}
\end{center}
\caption{\label{fig:proj}Projection of the fully filamented distribution onto the
$\tx-$axis for normalized kick amplitudes of $\theta/\sqrt{\eps/\beta}$ equal to
0,2,\dots,10 and $\eps=1\,$m\,rad.}
\end{figure}
%..............................................
In Fig.~\ref{fig:dist} we show the kicked distribution in normalized phase space after 
full filamentation. The kick amplitude used in this plot was $\theta=6\sqrt{\eps/\beta}$
or six times the angular divergence at the location where the kick is applied. We 
observe that the distribution is rotationally invariant and has a annular shape with
a hole in the center, which is what we intuitively expect. The distribution in
the transverse coordinate $\tx$ is given by projecting the two-dimensional distribution 
shown in Fig.~\ref{fig:dist} onto the spatial axis, which is equivalent to integrating over
the angle variable $\tx'.$ We show the result of the projection for different kick
amplitudes in Fig.~\ref{fig:proj}. For $\theta=0$ we recover the Gaussian with unit
width. For increasing kick amplitude we see that the peak value of the distribution 
gets smaller and there are two separate peaks with a reduced particle density 
in-between.
\par
%..............................................
\begin{figure}[tb]
\begin{center}
\epsfig{file=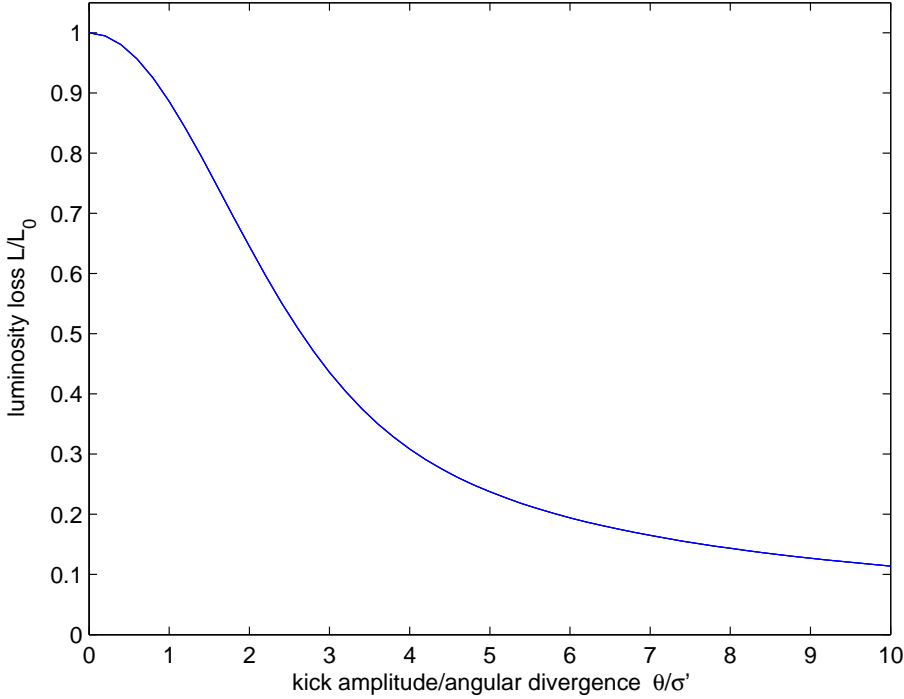,width=0.95\textwidth}
\end{center}
\caption{\label{fig:lumi}The relative luminosity loss as a function of the
kick amplitude. Note that the loss is calculated as the geometric overlap only.}
\end{figure}
%..............................................
In order to calculate the luminosity we need to determine the transverse spatial 
distribution at the interaction point, which is given by the the projection
of the distribution in normalized phase space, as shown in Fig.~\ref{fig:proj},
mapped back into real space. But this is accomplished by multiplying the
$\tx$ axis by $\sqrt{\beta_0}$ where $\beta_0$ is the beta function at the 
interaction point. But since this implies rescaling the axes we might as 
well calculate the relative luminosity loss by using normalized phase space
coordinates throughout.
\par
We estimate the detrimental effect of the kicks on the luminosity by calculating 
the reduction in geometric overlap of the kicked beam with the counter-propagating 
beam. This approximation is valid as long as no strong beam-beam enhancement due
to strong focusing of one beam on the other which results in a strong pinch
effect is present. Here the geometric overlap may serve as a qualitative measure
of how significant the luminosity reduction is. In a parameter regime with strong 
pinch effect, it can be stronger.
\par
In the present discussion we assume that the beam line is uncoupled the kick only 
affects the beam size in one plane and leaves the other unaffected. Thus we can 
calculate the luminosity loss due to fully filamented kicks by calculating the 
following integral
\begin{equation}
{\cal L}(\theta) = \frac{1}{\sqrt{2\pi\eps}}\int_{-\infty}^{\infty} e^{-\tx^2/2\eps}
\left[\frac{1}{2\pi\eps} \int_{-\infty}^{\infty} 
\psi_f\left(\sqrt{(\tx^2+\tx^{\prime 2})/\eps}\right)
 d\tx'\right] d\tx
\end{equation}
where $\psi_f$ is given by eq.~\ref{eq:psif} with the action variable $J$ substituted
by $J=(\tx^2+\tx^{\prime 2})/2$ such that the square bracket is the projection of the
filamented distribution onto the space coordinate. The second integral over $\tx$,
weighted with the spatial distribution of the counter-propagating beam then results
in the luminosity loss. This integral we evaluated numerically.
\par
In Fig.~\ref{fig:lumi} we show the luminosity, normalized to 
the value ${\cal L}_0$ at kick angle $\theta=0$ as a function of the kick angle in
units of the angular divergence at the location where the kick occurs. We find that
small kick angles of magnitude less than one sigma causes a luminosity reduction
on the order of 10\,\%. We also note that already for a kick of two-sigma magnitude 
almost halves the luminosity and are clearly very detrimental. We also observe that
even for very large kicks of ten sigma amplitude the filamentation causes some of 
the particles to be near the center of the beam line where they can interact with
the counter-propagating beam, albeit at  a significant lower rate. The luminosity at 
ten sigma is only about 10\,\% of the value when beams of equal size meet.
\par
We now turn to the analysis of the effect of a betatron mismatch, for instance due
to quadrupolar errors, on the luminosity.
%
%%%%%%%%%%%%%%%%%%%%%%%%%%%%%%%%%%%%%%%%%%%%%%%%%%%%%%%%%%%%%%%%%%%%%%%
%
\section{Betatron Mismatch}
In the case of a betatron mismatch the incoming beam has Twiss parameters different 
from those of the matched beam. This implies that the incoming distribution is not
rotationally invariant and filamentation will cause the beam size in general to  
grow from the initial one. We start by simply considering the emittance growth. This
is easily done by calculating the final sigma matrix 
\begin{equation}
\sigma_f=R \sigma R^t =
\left({\cal A}{\cal O} {\cal A}^{-1}\right) 
\eps \left(\bar{\cal A}\bar{\cal A}^t\right)
\left({\cal A}{\cal O} {\cal A}^{-1}\right)^t\ .
\end{equation}
Observe that the transfer matrix $R$ is based on the matched Twiss parameters through
the matrix $\cal A$ whereas the incoming beam is based on different Twiss parameters
$\bar\beta$ and $\bar\alpha$ that define the matrix $\bar{\cal A}$. Explicitely
multiplying out the matrices we find
\begin{equation}
\sigma_f = \eps{\cal A} {\cal O} 
\left(\begin{array}{cc}
\frac{\bbeta}{\beta} & \alpha\frac{\bbeta}{\beta}-\balpha\\
\alpha\frac{\bbeta}{\beta}-\balpha &
\alpha^2\frac{\bbeta}{\beta}-2\alpha\balpha+\bgamma\beta
\end{array}\right)
{\cal O}^t \left({\cal A}^{-1}\right)^t \ .
\end{equation}
Again, performing the multiplication with the matrices $\cal O$ results in a
$2\times 2$ matrix that contains terms with cosine and sine of the phases 
$\phi=\mu$ and filamentation corresponds to averaging over these phases
in the range $0$ to $2\pi$ causes the terms with $\cos^2\phi$ and $\sin^2\phi$
to be replaced by their average of 1/2 and terms $\sin\phi\cos\phi$ to be
replaced by zero. These simple but lengthy calculations finally lead to the 
final sigma matrix $\sigma_f$ averaged over the phases
\begin{equation}\label{eq:fil}
\langle\sigma_f\rangle_{\phi} = \eps {\cal A} {\cal A}^t B_{mag} 
\end{equation}
with the conventional definition of the emittance growth factor due to
betatron mismatch $B_{mag}$ given by~\cite{DECKER}
\begin{equation}\label{eq:bmag}
B_{mag} = \frac{1}{2}\left[\frac{\bbeta}{\beta}+\alpha^2\frac{\bbeta}{\beta}-
                         2\alpha\balpha+\bgamma\beta\right]
  = \frac{1}{2}\left[\left(\frac{\bbeta}{\beta} + \frac{\beta}{\bbeta}\right)
   +\beta\bbeta\left(\frac{\balpha}{\bbeta}-\frac{\alpha}{\beta}\right)^2\right]\ .
\end{equation}
Equation~\ref{eq:fil} shows that the final beam has a sigma matrix given by the
Twiss parameters of the matched beam, but an emittance that is given by the 
initial emittance $\eps$ multiplied by the growth factor $B_{mag}.$ For a matched
beam $ B_{mag}$ is unity. 
\par
%..............................................
\begin{figure}[tb]
\begin{center}
\epsfig{file=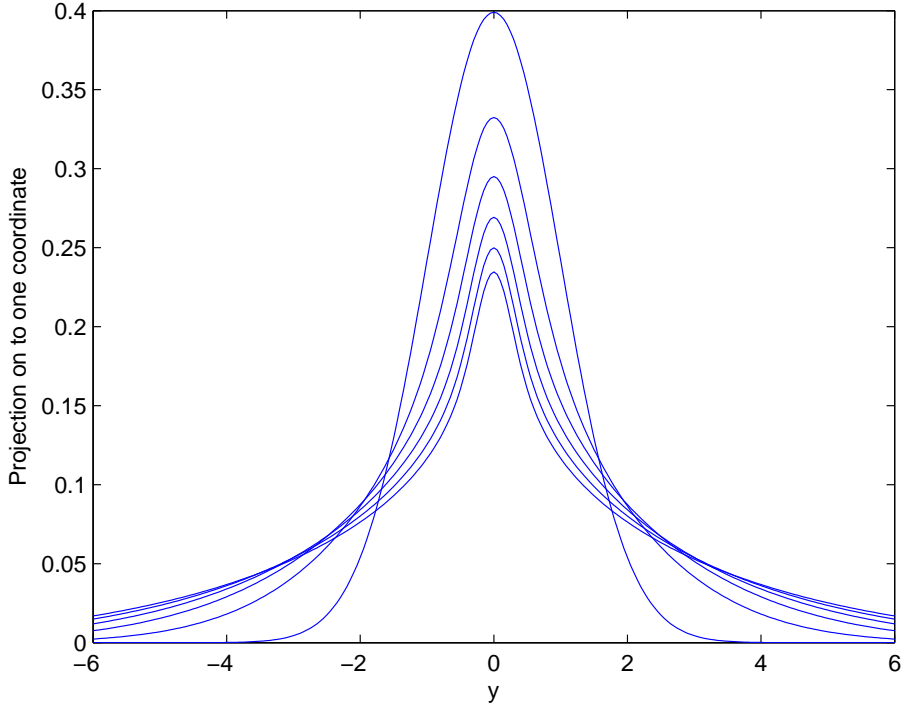,width=0.95\textwidth}
\end{center}
\caption{\label{fig:bbdist}Transverse distribution after filamentation of a 
mismatched beam characterized by $B_{mag}= 1,3,\dots,11$.}
\end{figure}
%..............................................
The discussion so far only considered the sigma matrix and consequently only 
rms properties of the beam distribution. We will drop that restriction now and
calculate the distribution after filamentation. To this end we consider the 
initial mismatched beam distribution given by 
\begin{eqnarray}
\psi(x,x') &=& \frac{1}{2\pi\eps}\exp\left[-\frac{1}{2}
 (x, x') \bar\sigma^{-1} \left(\begin{array}{c} x\\ x'\end{array}\right)\right]\\
&=& \frac{1}{2\pi\eps}\exp\left[-\frac{1}{2\eps}
  (x, x') \left(\bar{\cal A}\bar{\cal A}^t\right)^{-1}
\left(\begin{array}{c} x\\ x'\end{array}\right)\right]\nonumber
\end{eqnarray}
with the initial beam matrix $\bar\sigma$ given by eq.~\ref{eq:sigma}. We now
apply the same method as in the previous section and introduce variables of
normalized phase space through eq.~\ref{eq:nps}. Since the Jacobian of that
transformation is unity we can simply re-express the variables in the exponent
and arrive at the distribution function in the variables $\tx$ and $\tx'$
\begin{equation}
\psi(\tx,\tx') = \frac{1}{2\pi\eps} \exp\left[-\frac{1}{2\eps}(\tx, \tx')
{\cal A}^t\left(\bcA^t\right)^{-1} \bcA^{-1} {\cal A} 
\left(\begin{array}{c} \tx\\ \tx'\end{array}\right)\right]\ .
\end{equation}
The product of the four matrices ${\cal A}$ in the exponent leads to
\begin{equation}
{\cal A}^t\left(\bcA^t\right)^{-1} \bcA^{-1} {\cal A} 
=\left(\begin{array}{cc} c & b \\ b & a\end{array}\right) =
\left(\begin{array}{cc}
\frac{\beta}{\bbeta}+\balpha^2\frac{\beta}{\bbeta}-2\alpha\balpha+\alpha^2\frac{\bbeta}{\beta} &
\balpha -\alpha\frac{\bbeta}{\beta} \\
\balpha -\alpha\frac{\bbeta}{\beta} & \frac{\bbeta}{\beta}
\end{array}\right)
\end{equation}
and the distribution function can be expressed as
\begin{equation}
\psi(\tx,\tx') = \frac{1}{2\pi\eps} 
\exp\left[-\frac{c\tx^2 + 2 b \tx\tx' + a \tx^{\prime 2}}{2\eps}\right]\ .
\end{equation}
Rewriting the distribution equation in terms of action-angle variables using
eq.~\ref{eq:aa} we obtain
\begin{equation}
\psi(J,\phi) = \frac{1}{2\pi\eps}
\exp\left[-\frac{J\left\{ (c+a) + (c-a)\cos(2\phi)+2b\sin(2\phi)\right\}}{2\eps}\right] 
\end{equation}
which can be simplified to
\begin{equation}
\psi(J,\phi) = \frac{1}{2\pi\eps}
\exp\left[-\frac{J}{2\eps} 
\left\{ (c+a) + \sqrt{(c-a)^2+4b^2}\cos(2\phi-\delta)\right\}\right]
\end{equation}
by introducing
\begin{equation}
\tan\delta = \frac{2b}{c-a}
\end{equation}
which has the advantage that only a single trigonometric function appears in the
exponent and this aids evaluating the average over the angle variable we need to
perform in order to calculate the fully filamented distribution  
\begin{eqnarray}
\psi_f(J) &=& \frac{1}{2\pi} \int_0^{2\pi} \psi(J,\phi) d\phi\nonumber\\
&=& \frac{e^{-J(a+c)/2\eps}}{2\pi\eps} \frac{1}{2\pi}\int_0^{2\pi}
\exp\left[ -\frac{J\sqrt{(c-a)^2+4b^2}}{2\eps}\cos(2\phi-\delta)\right]
\nonumber\\
&=& \frac{e^{-J(a+c)/2\eps}}{2\pi\eps} I_0\left(\frac{J\sqrt{(c-a)^2+4b^2}}{2\eps}\right)
\end{eqnarray}
where we have, again, used the integral representation of the modified Bessel function
$I_0$ from ref.~\cite{ABRASTE}. In fact, we can express the factors $a,b,c$ to more
commonly used quantities, by observing that
\begin{eqnarray}
\frac{a+c}{2} &=& \frac{1}{2}\left[  \frac{\bbeta}{\beta} +
\frac{\beta}{\bbeta}+\balpha^2\frac{\beta}{\bbeta}-2\alpha\balpha+\alpha^2\frac{\bbeta}{\beta}
\right] = B_{mag} \\
(c-a)^2+4b^2 &=& (a+c)^2 - 4(ac-b^2) = 4 ( B_{mag}^2 - 1) \nonumber
\end{eqnarray}
where we note that $ac-b^2=1$ is the determinant of the four matrices $\cal A$ each 
having unity determinant. This allows us to write the filamented distribution as
\begin{equation}
\psi_f(J) = \frac{1}{2\pi\eps}e^{-B_{mag}J/\eps} I_0\left(\frac{J}{\eps}\sqrt{B_{mag}^2-1}\right)
\end{equation}
which coincides with the expression derived in ref.~\cite{TOR}.
\par
%..............................................
\begin{figure}[tb]
\begin{center}
\epsfig{file=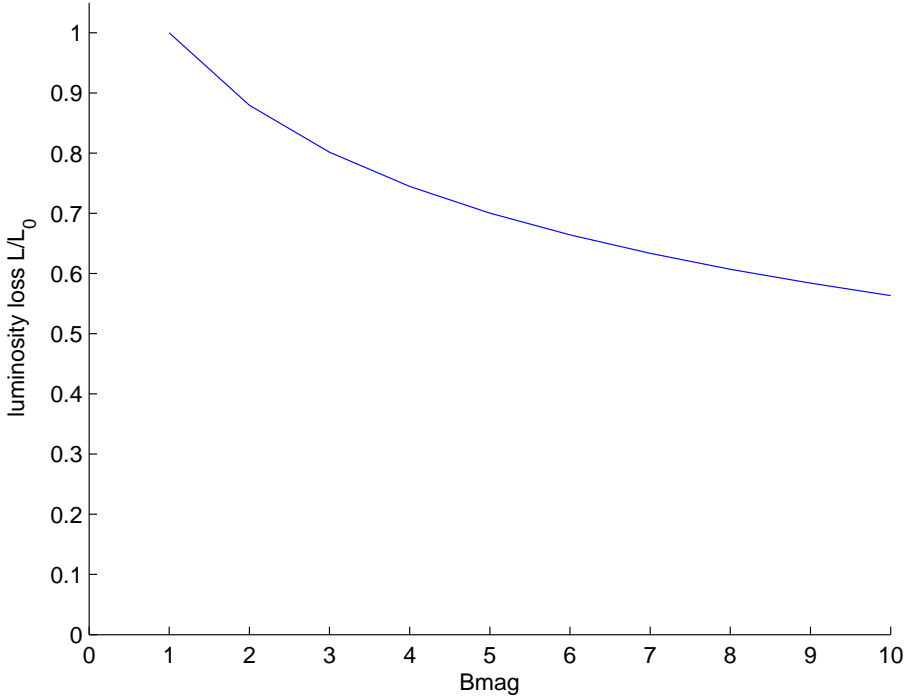,width=0.95\textwidth}
\end{center}
\caption{\label{fig:bblumi}The relative reduction of the luminosity as a function
of the betatron mismatch parameter $B_{mag}.$}
\end{figure}
%..............................................
Similar to the previous section we now calculate the spatial distribution 
by re-introducing normalized phase space variable through 
$J=(\tx^2+\tx^{\prime 2})/2$ in the previous equation and integrate over the 
angle variable $\tx'$. We do this integral numerically for different mismatch 
parameter $B_{mag}$ and show the resulting distributions in Fig.~\ref{fig:bbdist}.
There we note that the peak intensity in the center drops with increasing
$B_{mag}$ and that the tails are increasingly populated. 
\par
Since there are less particles in the tail we also expect a reduced luminosity 
that we calculate in the same way as before by averaging the distribution in 
Fig.~\ref{fig:bbdist} with that of the counter-propagating beam that is assumed 
matched and show the result in Fig.~\ref{fig:bblumi}. A mismatch parameter
$B_{mag}=2$ approximately causes a 10\,\% loss of luminosity.
\par
It is instructive to relate the mismatch parameter to an integrated quadrupole 
gradient error that is characterized by its focal length $f.$ We calculate
the beam matrix after the quadrupole error $\bar\sigma$ and compare it to the 
beam matrix $\sigma$ that had not experienced the quadrupole error. For
$\bar\sigma$ we have
\begin{equation}
\eps\left(\begin{array}{cc} \bbeta &-\balpha \\ -\balpha & \bgamma\end{array}\right) 
= \left(\begin{array}{cc} 1 & 0 \\ -1/f & 1\end{array}\right) 
\eps\left(\begin{array}{cc} \beta &-\alpha \\ -\alpha & \gamma\end{array}\right)
 \left(\begin{array}{cc} 1 & -1/f \\ 0 & 1\end{array}\right) \ .
\end{equation}  
Evaluating the matrix multiplications we find
\begin{equation}
\bbeta= \beta \qquad\mathrm{and}\qquad \balpha = \alpha+\frac{\beta}{f}
\end{equation}
and inserting $\bbeta$ and $\balpha$ into the expression for $B_{mag}$ in eq.~\ref{eq:bmag}
we find that a localized quadrupole error causes a beta mismatch of
\begin{equation}
B_{mag}=1+\frac{\beta^2}{2f^2}\ .
\end{equation}
If an integrated quadrupole error $k_2l=1/f$ has such a strength that the focal length
is half the beta function at that location, it causes a doubling of the emittance after
filamentation, but that is a rather strong quadrupole error. 
\section{Mismatch due to a Skew Quadrupole}
We now investigate the filamentation from a thin skew quadrupolar error and
derive the loss of luminosity as characterized by the geometric overlap, as before.
We start by only calculating the emittance growth due to filamentation at the 
final point, the interaction point, of the beam line that is caused by the 
skew quadrupole.
\par
The transfer matrix $S$ of a thin skew quadrupole with focal length $f$ is given by
\begin{equation}
S=\left(\begin{array}{cccc}
    1 & 0 &  0  & 0 \\
    0 & 1 & 1/f & 0\\
    0 & 0 &  1  & 0\\
   1/f& 0 &  0  & 1
\end{array}\right)\ .
\end{equation}
We now assume that an uncoupled beam characterized by the $4\times 4$ beam matrix
\begin{equation}\label{eq:sigmaini}
\bar\sigma=\left(\begin{array}{cc}
\sigma_x &   0_2\\
  0_2    & \sigma_y
\end{array}\right)\ .
\end{equation}
where $0_2$ is the $2\times 2$ matrix containing zeros, only and $\sigma_x$ and 
$\sigma_y$ are $2\times 2$ beam matrices as given in eq.~\ref{eq:sigma}. Furthermore 
we assume that it describes the matched beam. Propagating the matched beam 
$\bar\sigma$ through the skew quadrupoles we arrive at
\begin{equation}\label{eq:sigmahat}
\hat\sigma=S\bar\sigma S^t = 
\left(\begin{array}{cc} 1_2 & Q \\ Q & 1_2\end{array}\right)
\left(\begin{array}{cc} \sigma_x & 0_2 \\ 0_2 & \sigma_y\end{array}\right)
\left(\begin{array}{cc} 1_2 & Q^t \\ Q^t & 1_2\end{array}\right)
\end{equation}
where $1_2$ is the two-dimensional unit matrix, $0_2$ the $2\times2$ 
zero-matrix, and $Q$ is given by
\begin{equation}\label{eq:Q}
Q=
\left(\begin{array}{cc} 0   & 0 \\ 1/f & 0 \end{array}\right)\ .
\end{equation}
Evaluating the matrix products we find
\begin{equation}
\hat\sigma=
\left(\begin{array}{cc}
\sigma_x + Q\sigma_y Q^t & \sigma_x Q^t + Q\sigma_y \\
Q\sigma_x+\sigma_yQ^t & \sigma_y+Q\sigma_x Q^t
\end{array}\right)\ .
\end{equation}
This expression is the sigma matrix of a matched beam after it had passed 
the skew quadrupole. In order to find the beam matrix at the interaction point
we need to propagate it with the $4\times 4$ transfer matrix $\bar R$ given
in terms of the matched beta functions and phase advance as given in
eq.~\ref{eq:Rnps} for the horizontal and vertical plane
\begin{equation}
\bar R =
\left(\begin{array}{cc}
{\cal A}_x {\cal O}_x {\cal A}_x^{-1} &   0_2 \\
0_2 & {\cal A}_y {\cal O}_y {\cal A}_y^{-1} \\
\end{array}\right)\ .
\end{equation}
The final $4 \times 4$ beam matrix at the interaction point $\tilde\sigma$ is then
given by 
\begin{equation}
\tilde\sigma=R S\bar\sigma S^t R^t = 
\left(\begin{array}{cc}
    \tilde\sigma_x & \mathrm{linear\ in\ {\cal O}_x\ and\ {\cal O}_y}\\
    \mathrm{linear\ in\ {\cal O}_x\ and\ {\cal O}_y} & \tilde\sigma_y
\end{array}\right)
\end{equation}
with
\begin{eqnarray}
\tilde\sigma_x &=& \sigma_x +\left({\cal A}_x {\cal O}_x {\cal A}_x^{-1}\right) 
      Q\sigma_y Q^t \left({\cal A}_x {\cal O}_x {\cal A}_x^{-1}\right)\nonumber\\
\tilde\sigma_y &=& \sigma_y +\left({\cal A}_y {\cal O}_y {\cal A}_y^{-1}\right) 
      Q\sigma_x Q^t \left({\cal A}_y {\cal O}_y {\cal A}_y^{-1}\right) \ .
\end{eqnarray}
Note that the off diagonal blocks in $\tilde\sigma$ are linear in the 
phase advance matrices ${\cal O}_x$ and  ${\cal O}_y$ and therefore vanish
when averaging over the phases. This implies that the beam after filamentation 
is uncoupled but the respective $2 \times 2$ sigma matrices are affected.
\par
The expressions for $\tilde\sigma_{x/y}$ can be evaluated in a straightforward, 
albeit tedious way and after averaging over the phases in the rotation matrices
${\cal O}_x$ and  ${\cal O}_y$ we find
\begin{equation}
\tilde\sigma=
\left(\begin{array}{cc}
\sigma_x +\frac{\eps_y}{2\eps_x}\frac{\beta_x\beta_y}{f^2}\sigma_x & 0_2 \\
0_2 & \sigma_y +\frac{\eps_x}{2\eps_y}\frac{\beta_x\beta_y}{f^2}\sigma_y
\end{array}\right)
\end{equation}
and after expressing the matched beam matrices $\sigma_x$ and $\sigma_y$ using
eq.~\ref{eq:sigma} we find the rms emittances $\tilde\eps_{x/y}$ at the 
interaction point to be
\begin{equation}\label{eq:skewemit}
\tilde\eps_x = \eps_x + \frac{\kappa^2}{2}\eps_y\quad\mathrm{and}\quad
\tilde\eps_y = \eps_y + \frac{\kappa^2}{2}\eps_x
\end{equation}
where we introduced the coupling factor $\kappa=\beta_x\beta_y/f^2.$ Remember
that the beta function appearing in the definition of $\kappa$ are those of the
skew quadrupole and we find that the emittance growth is proportional to both
beta-functions at the skew quadrupole and also proportional to inverse focal length,
the strength, of the skew quadrupole. Equation~\ref{eq:skewemit} also tells us
that the emittance of the orthogonal plane is projected into the plane. In this
way it is obvious that the smaller emittance, usually the vertical, is worse
affected by skew errors, because the larger horizontal emittance is added to it
proportional to the coupling strength $\kappa^2.$ 
\par
The discussion so far only took into account the averaged effect on the rms 
emittances, but not the shape of the beam distribution after filamentation. This
is what we address in the following paragraphs. We start by considering the 
matched beam distribution function just before the skew quadrupole
\begin{equation}
\psi(x,x',y,y') = \frac{1}{(2\pi)^2\eps_x\eps_y}
\exp\left[ -\frac{1}{2}\sum_{i=1}^4\sum_{j=1}^4\bar\sigma_{ij}^{-1}x_ix_j\right]
\end{equation}
where we introduce the notation $x_i$ for $i=1,..,4$ denotes $(x,x',y,y')$ for
convenience. Note that the inverse of the beam matrix from eq.~\ref{eq:sigmaini}
appears in the argument of the exponential function. 
\par
The task at hand is now to calculate the distribution function after the skew
quadrupole, introduce action-angle variables and then perform the phase averaging
in order to calculate the distribution after filamentation at the interaction
point. The inverse of the beam matrix before the skew quad is 
\begin{equation}
\bar\sigma^{-1} = 
\left(\begin{array}{cc}
\frac{1}{\eps_x} \left(\begin{array}{cc}\gamma_x&\alpha_x\\ \alpha_x&\beta_x\end{array}\right) &
0_2 \\ 0_2 &
\frac{1}{\eps_y} \left(\begin{array}{cc}\gamma_y & \alpha_y\\ \alpha_y&\beta_y\end{array}\right)
\end{array}\right)
\end{equation}
and we now use this to calculate the inverse of the beam matrix after the
skew quadrupole $\hat\sigma^{-1}$. Since $\hat\sigma$ is given by eq.~\ref{eq:sigmahat} 
we can easily invert it by writing
\begin{equation}
\hat\sigma^{-1}=(S^t)^{-1}\bar\sigma^{-1}S^{-1}=
\left(\begin{array}{cc} 1_2 & Q^t \\ Q^t & 1_2 \end{array}\right)^{-1} 
\bar\sigma^{-1}
\left(\begin{array}{cc} 1_2 & Q \\ Q & 1_2 \end{array}\right)^{-1} 
\end{equation}
and some extensive algebra, where we use the fact the $Q^2=0_2$ with $Q$ from
eq.~\ref{eq:Q} and also 
\begin{equation}
\left(\begin{array}{cc} 1_2 & Q \\ Q & 1_2 \end{array}\right)^{-1}
=
\left(\begin{array}{cc} 1_2 & -Q \\ -Q & 1_2 \end{array}\right)\ .
\end{equation}
we arrive at the inverse after the skew quadrupole
\begin{equation}\label{eq:sigmahatinv}
\hat\sigma^{-1}=
\left(\begin{array}{cc}
\sigma_x^{-1}+Q^t\sigma_y^{-1}Q & -\sigma_x^{-1} Q - Q^t\sigma_y^{-1} \\
-Q^t\sigma_x^{-1}- \sigma_y^{-1}Q & \sigma_y^{-1}+Q^t\sigma_x^{-1}Q
\end{array}\right)
\end{equation}
such that the distribution after the skew quadrupole now looks like
\begin{equation}
\psi(x,x',y,y') = \frac{1}{(2\pi)^2\eps_x\eps_y}
\exp\left[ -\frac{1}{2}\sum_{i=1}^4\sum_{j=1}^4\hat\sigma_{ij}^{-1}x_ix_j\right]\ .
\end{equation}
Essentially, only $\bar\sigma$ was changed to $\hat\sigma$ in the argument of the
exponent and $\hat\sigma$ is given by the previous equation.
\par
In order to propagate this distribution further down the assumed uncoupled linac we 
introduce normalized phase space coordinates in four dimensions by
\begin{equation}
\left(\begin{array}{c} x \\ x' \\ y \\ y'\end{array}\right)
=
\left(\begin{array}{cc} {\cal A}_x & 0_2 \\ 0_2 & {\cal A}_y\end{array}\right)
\left(\begin{array}{c} \tilde x \\ \tilde x' \\ \tilde y \\ \tilde y'\end{array}\right)
\end{equation}
where the ${\cal A}_{x/y}$ are defined in eq.~\ref{eq:nps}. If we want to express the 
argument of the exponential function in the variables of normalized phase space we 
need to calculate the inverse of the beam matrix $\tilde\sigma^{-1}$ in those variables
and find
\begin{equation}
\tilde\sigma^{-1}= 
\left(\begin{array}{cc} {\cal A}^t_x & 0_2 \\ 0_2 & {\cal A}^t_y\end{array}\right)
\hat\sigma^{-1}
\left(\begin{array}{cc} {\cal A}_x & 0_2 \\ 0_2 & {\cal A}_y\end{array}\right)
\end{equation}
with $\hat\sigma^{-1}$ given by eq.~\ref{eq:sigmahatinv}. Some more tedious algebra
yields
\begin{equation}
\tilde\sigma^{-1}= 
\left(\begin{array}{cc}
{\cal A}^t_x\sigma_x^{-1}{\cal A}_x + {\cal A}^t_x Q^t \sigma_y^{-1} Q  {\cal A}^t_x &
-{\cal A}^t_x\left[\sigma_x^{-1} Q + Q^t\sigma_y^{-1}\right]{\cal A}^t_y \\
-{\cal A}^t_y\left[Q^t\sigma_x^{-1} + \sigma_y^{-1}Q\right]{\cal A}^t_x &
{\cal A}^t_y\sigma_y^{-1}{\cal A}_y + {\cal A}^t_y Q^t \sigma_x^{-1} Q  {\cal A}^t_y 
\end{array}\right)\ .
\end{equation}
Now all the terms can be evaluated individually with the result 
\begin{eqnarray}
{\cal A}^t_x\sigma_x^{-1}{\cal A}_x &=& \frac{1}{\eps_x} 1_2\nonumber\\
{\cal A}^t_x Q^t \sigma_y^{-1} Q  {\cal A}^t_x &=& 
  \frac{1}{\eps_y}\left(\begin{array}{cc}\kappa^2 & 0 \\ 0 & 0 \end{array}\right)
\nonumber\\
{\cal A}^t_x\sigma_x^{-1} Q{\cal A}_y &=& 
  \frac{1}{\eps_x}\left(\begin{array}{cc}0 & 0 \\ \kappa & 0 \end{array}\right)
\\
{\cal A}^t_xQ^t\sigma_y^{-1} {\cal A}_y &=&
  \frac{1}{\eps_y}\left(\begin{array}{cc}0 & \kappa\\ 0 & 0 \end{array}\right)
\nonumber\\
{\cal A}^t_yQ^t\sigma_x^{-1} {\cal A}_x &=&
  \frac{1}{\eps_x}\left(\begin{array}{cc}0 & \kappa\\ 0 & 0 \end{array}\right)
\nonumber\\
{\cal A}^t_y\sigma_y^{-1} Q {\cal A}_x &=&
  \frac{1}{\eps_y}\left(\begin{array}{cc}0 & 0 \\ \kappa & 0 \end{array}\right)
\nonumber
\end{eqnarray}
where we use the definition of $\kappa^2=\beta_x\beta_y/f^2$ from above. Inserting 
all terms into the expression for $\tilde\sigma^{-1}$ we obtain
\begin{equation}
\tilde\sigma^{-1}= 
\left(\begin{array}{cccc}
    \frac{1}{\eps_x}+\frac{\kappa^2}{\eps_y} & 0  & 0 & \frac{1}{\eps_y}\\
    0 & -\frac{\kappa}{\eps_x}  & -\frac{\kappa}{\eps_x} & 0 \\
    0 & -\frac{\kappa}{\eps_x} &  \frac{1}{\eps_y}+\frac{\kappa^2}{\eps_x} & 0 \\ 
    -\frac{\kappa}{\eps_y} & 0 & 0 & \frac{1}{\eps_y}
\end{array}\right)\ .
\end{equation}
Note that in the absence of coupling $\kappa=0$ the matrix reverts to the 
matrix that contains the inverse emittances on the diagonal. 
\par
In order to express the distribution function in action-angle variables $J$ and $\phi$
we change the variables according to
\begin{equation}\label{eq:npsaa}
\tilde z 
= 
\left(\begin{array}{c} \tilde x \\ \tilde x' \\ \tilde y \\ \tilde y'\end{array}\right)
=
\left(\begin{array}{c} 
\sqrt{2J_x}\cos\phi_x \\ \sqrt{2J_x}\sin\phi_x\\ \sqrt{2J_y}\cos\phi_y \\ \sqrt{2J_y}\sin\phi_y 
\end{array}\right)
\end{equation}
which lets us express the argument of the exponential function in the distribution function
(apart from the factor 1/2) in the form
\begin{eqnarray}
\tilde z^t \tilde\sigma^{-1} \tilde z &=&
\frac{2J_x}{\eps_x} + \frac{2J_x\kappa^2}{\eps_y}\cos^2\phi_x
-\frac{2\kappa}{\eps_y}\sqrt{2J_x}\sqrt{2J_y}\cos\phi_x\sin\phi_y\nonumber\\
& &+ \frac{2J_y}{\eps_y} + \frac{2J_y\kappa^2}{\eps_x}\cos^2\phi_y
-\frac{2\kappa}{\eps_x}\sqrt{2J_x}\sqrt{2J_y}\cos\phi_y\sin\phi_x 
\end{eqnarray}
and the distribution function in action angle variables can finally be written as
\begin{eqnarray}\label{eq:fullpsi}
\psi(J_x,\phi_x,J_y,\phi_y) &=& \frac{1}{(2\pi)^2\eps_x\eps_y} 
\exp\left[-\frac{J_x}{\eps_x}- \frac{J_y}{\eps_y} - \frac{\kappa^2 J_x}{\eps_y}\cos^2\phi_x
-\frac{\kappa^2 J_y}{\eps_x}\cos^2\phi_y\right]\nonumber\\
& &\times \exp\left[\frac{\kappa}{\eps_y}\sqrt{2J_x}\sqrt{2J_y}\cos\phi_x\sin\phi_y
\right. \nonumber\\
& & \qquad\quad\left. +\frac{\kappa}{\eps_x}\sqrt{2J_x}\sqrt{2J_y}\cos\phi_y\sin\phi_x\right]\ .
\end{eqnarray}
The effect of filamentation is, similar to what we did in the simple cases in previous sections, 
given by the phase average over the angle variables $\phi_x$ and $\phi_y$ and we calculate
\begin{equation}
\langle\psi(J_x,J_y)\rangle_{\phi_x,\phi_y}=
\frac{1}{(2\pi)^2}\int_{-\pi}^{\pi} \int_{-\pi}^{\pi} \psi(J_x,\phi_x,J_y,\phi_y)\, d\phi_xd\phi_y
\end{equation}
which, unfortunately, we were unable to solve analytically. It is, however, straightforward
to evaluate the integral numerically on a grid of values for $J_x$ and $J_y$ for given 
parameters $\eps_x,\eps_y,$ and $\kappa.$ We chose to normalize the variables $J_x, J_y,$
and $\eps_x$ to the vertical emittance $\eps_y$ which we consequently set to unity. 
\par
In order to calculate the luminosity degradation as a function of the coupling $\kappa$
for different emittance ratios $\eps_x/\eps_y=10,20,\dots,50$ we use the phase-averaged 
distribution that is defined on the grid, and express the action
variables $J_x$ and $J_y$ by 'new' variables $\tx,\tx',\ty,\ty'$ in normalized phase space
$J_x=(\tx^2+\tx'^2)/2$ and the corresponding expression for the vertical plane. Note,
that this is the normalized phase space {\em after} filamentation which is different
from the phase space used in eq.~\ref{eq:npsaa} despite our use of the same symbols for 
the variables. 
\par
%..............................................
\begin{figure}[tb]
\begin{center}
\epsfig{file=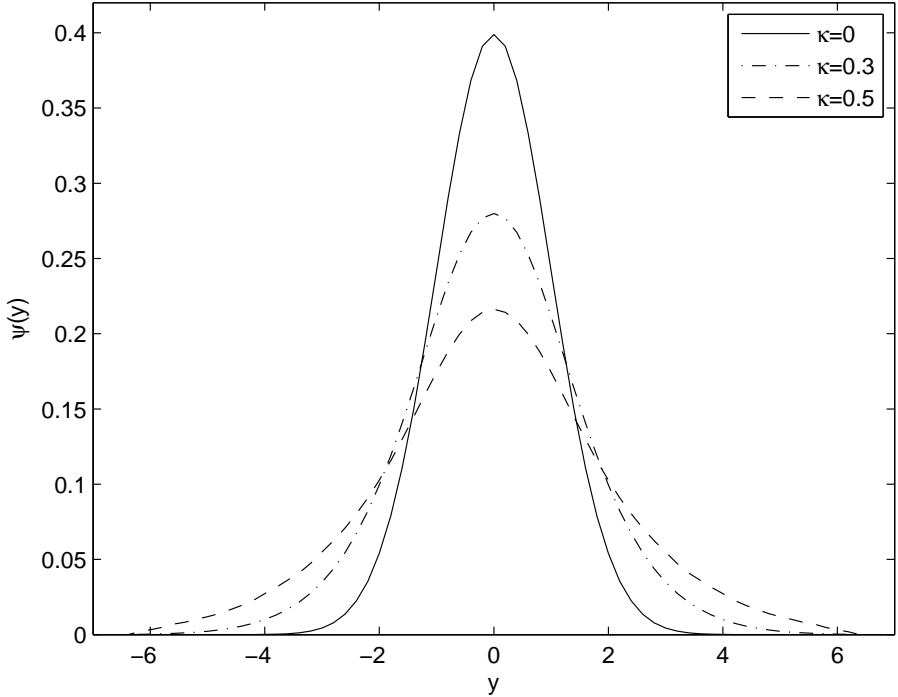,width=0.95\textwidth}
\end{center}
\caption{\label{fig:distris}The normalized vertical distribution for $\eps_x/\eps_y=30$ 
and $\kappa=0.0, 0.3$ and $0.5.$}
\end{figure}
%..............................................
The resulting distribution function is now a function of the 'new' normalized phase space
variables and the spatial distribution can be calculated by numerically by integrating
over the angle variables $\tx'$ and $\ty'$ which determine the action variables 
$J_x=(\tx^2+\tx'^2)/2$ and $J_y=(\ty^2+\ty'^2)/2$ that determine the value of the
distribution function $\psi$ by interpolating on the grid. This procedure results
in a distribution function $\psi_{\kappa}(\tx,\ty)$ which describes the spatial 
distribution of the beam, expressed in variables of the 'new' normalized phase space
after filamentation. It is
instructive to display the vertical distribution, which is obtained by numerically
integrating once more over $\tx$ which is shown if Fig.~\ref{fig:distris} for an emittance
ratio $\eps_x/\eps_y=30$ and for $\kappa=0, 0.2, 0.5.$ We observe the expected widening
of the distribution, which comes from projecting the large horizontal emittance onto
the vertical plane.
\par
%..............................................
\begin{figure}[tb]
\begin{center}
\epsfig{file=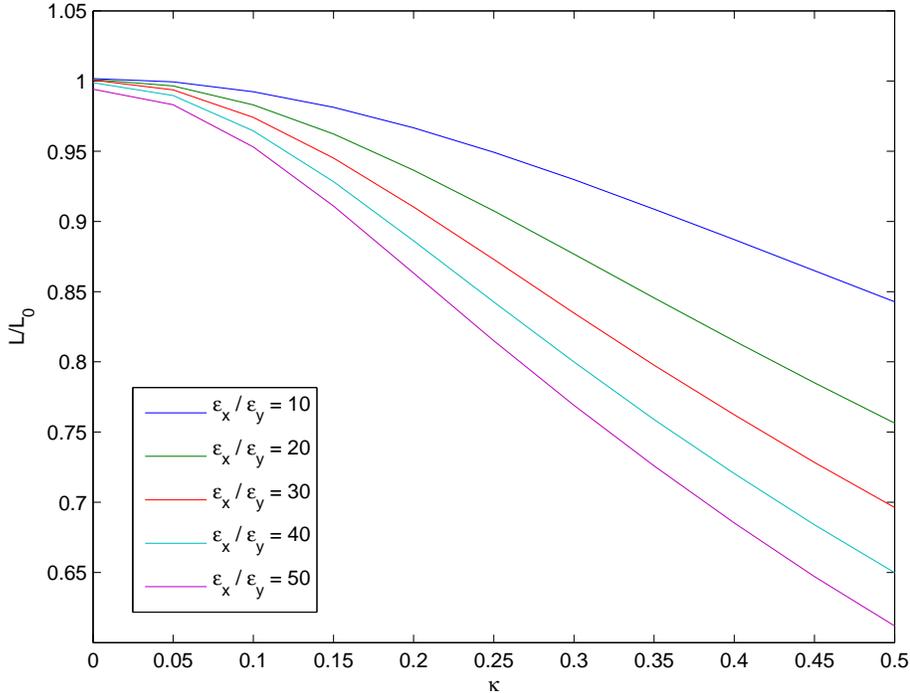,width=0.95\textwidth}
\end{center}
\caption{\label{fig:sumsum}The luminosity loss as a function of the coupling parameter 
$\kappa$ for different emittance ratios $\eps_x/\eps_y.$}
\end{figure}
%..............................................
The geometric luminosity is proportional to the spatial overlap integral of the distribution
with a given value of $\kappa$ with the distribution of the counter-propagating beam, that
we assume to be the unperturbed on, which corresponds to $\kappa=0.$ 
\begin{equation}
{\cal L} \propto \int_{-\infty}^{\infty} \int_{-\infty}^{\infty} 
    \psi_0(\tx,\ty)\psi_{\kappa}(\tx,\ty)d\tx d\ty\ .
\end{equation}
In the following we will characterize the loss of geometric luminosity by normalizing
with respect to the overlap integral of two unperturbed distributions $\psi_0(\tx,\ty).$
The unperturbed distributions are pure Gaussians, characterized by $\eps_x$ and $\eps_y$,
and the integral can be evaluated analytically with the result ${\cal L}_0\propto
1/4\pi\eps_x\eps_y.$ In Fig.~\ref{fig:sumsum} we show the relative geometric luminosity 
loss ${\cal L}/{\cal L}_0$ as a function of the coupling perturbation $\kappa$ for 
emittance ratios $\eps_x/\eps_y=10,20,\dots,50.$ We find, not surprisingly, that the
luminosity decreases with increasing coupling and that the effect is worse for larger
emittance ratios $\eps_x/\eps_y.$ A coupling factor of $\kappa=0.1$ results in reductions 
of 0.9\,\%, 1.8\,\%, 2.7\,\%, 3.4\,\%, and 4.1\,\%, respectively for the different
emittance ratios.
\par
\begin{figure}[tb]
\begin{center}
\epsfig{file=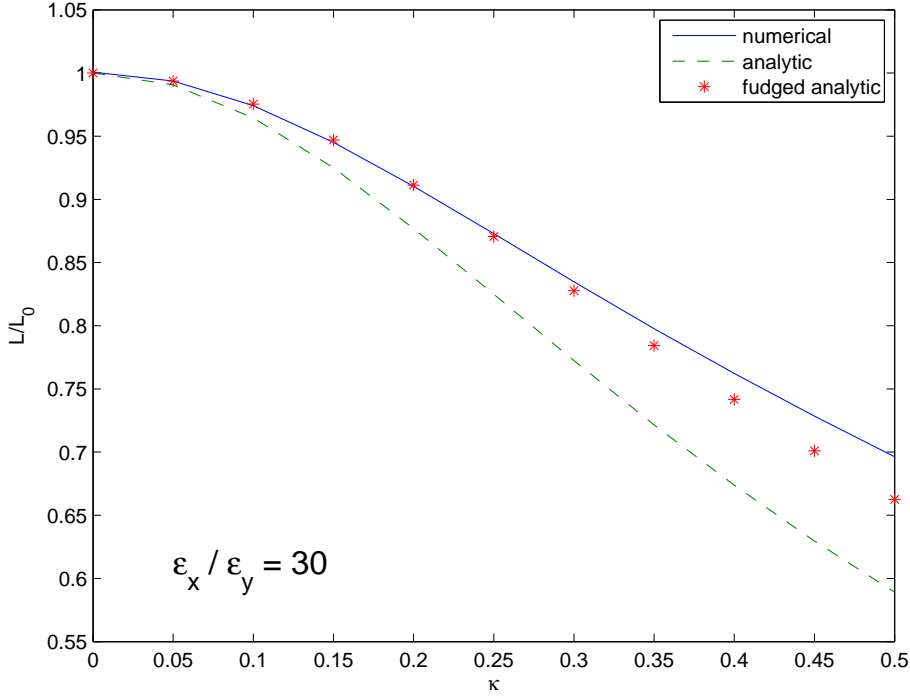,width=0.95\textwidth}
\end{center}
\caption{\label{fig:comp}Comparison of the approximations for the luminosity loss 
given in eq.~\ref{eq:lumiapprox} (dashed) and eq.~\ref{eq:lumireal} (asterisks) with the
numerically evaluated values (solid).}
\end{figure}

It is interesting to compare the luminosity loss ${\cal L}$ calculated using the proper 
distribution with the luminosity ${\cal L}^g$ derived from convoluting a Gaussian distribution 
whose rms is given by eq.~\ref{eq:skewemit} for the corresponding value of $\kappa.$ The 
latter can be calculated by averaging two Gaussians and we obtain
\begin{equation}\label{eq:lumiapprox}
{\cal L}^g/{\cal L}_0 = \frac{1}{\sqrt{1+(1/2)\kappa^2\eps_y/\eps_x}}
  \frac{1}{\sqrt{1+(1/2)\kappa^2\eps_x/\eps_y}}
\end{equation}
In Fig.~\ref{fig:comp}
we display the proper numerical luminosity loss for $\eps_x/\eps_y=30$ as a blue solid
line and the luminosity emittance loss from eq.~\ref{eq:lumiapprox} as a green dashed line.
We find that the emittance loss based on eq.~\ref{eq:lumiapprox} overestimates the loss.
By empirically tuning the parameter 1/2 in front of the $\kappa^2$ terms to 0.34 results
in the red asterisks in Fig.~\ref{fig:comp} and yields a much improved approximation of 
the proper numerical result shown as the solid line.
\par
%..............................................
\begin{figure}[tb]
\begin{center}
\epsfig{file=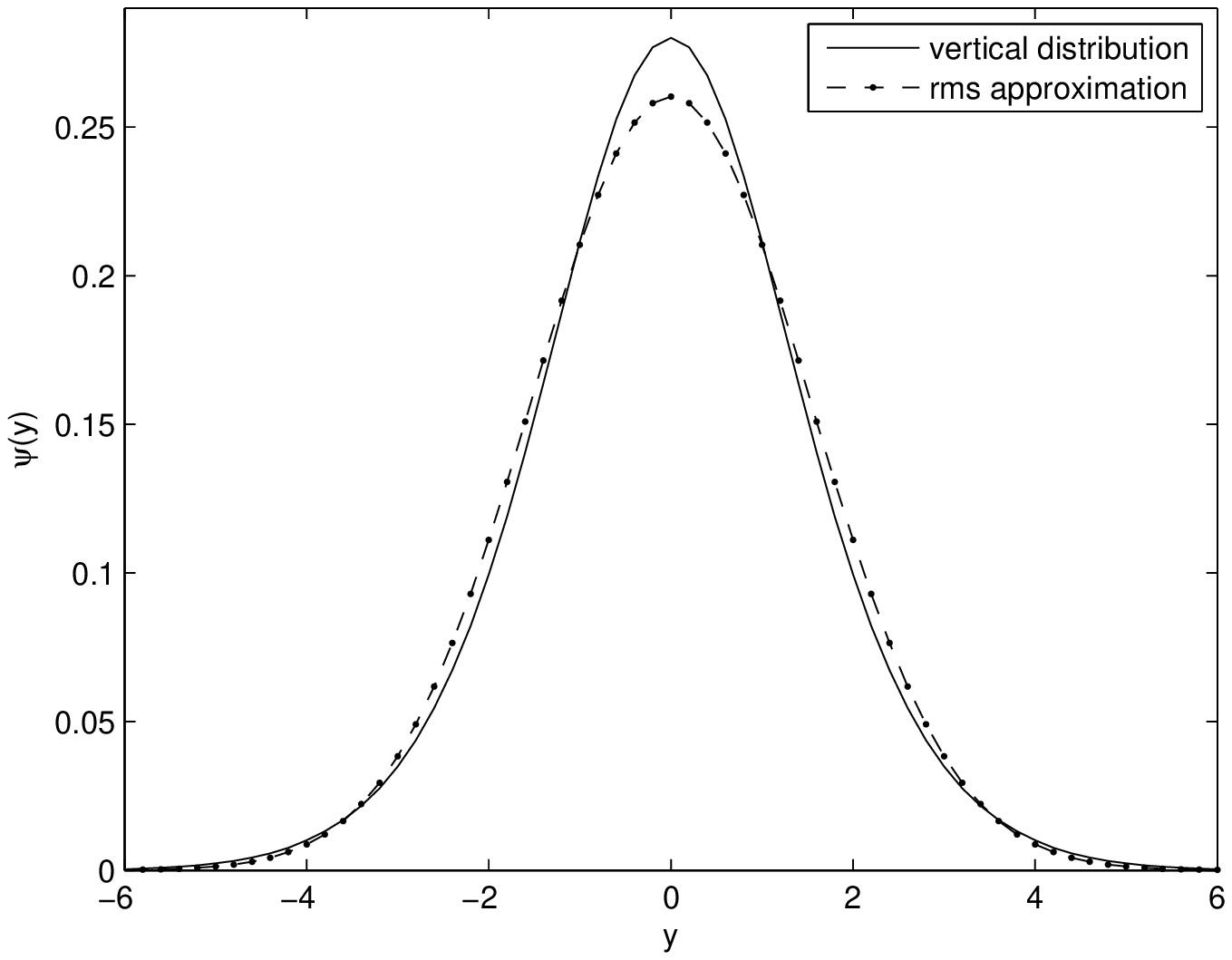,width=0.45\textwidth}
\epsfig{file=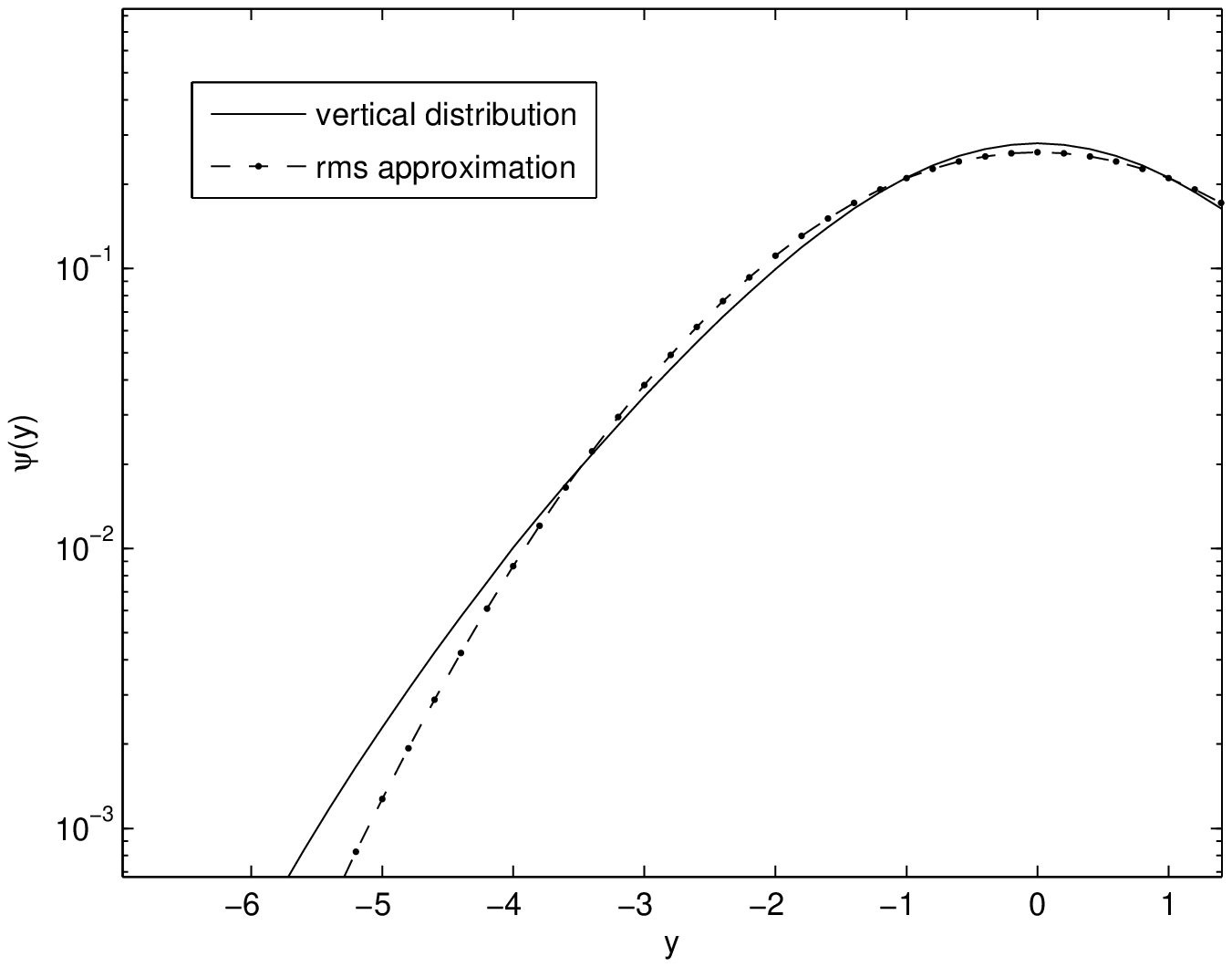,width=0.45\textwidth}
\end{center}
\caption{\label{fig:e30}The vertical distribution for $\eps_x/\eps_y=30$ and 
$\kappa=0.3$ is shown as a solid line and the approximation based on eq.~\ref{eq:skewemit} 
is shown as a dashed line. From the right plot, which shows the same data on a logarithmic 
scale, we see that the tail of the solid is larger than the dashed distribution. This 
leads to an overestimate of the rms value on which eq.~\ref{eq:skewemit} is based.}
\end{figure}

%..............................................
The reason for the improved approximation by the empirically found parameter can be
understood by plotting the spatial projection onto the vertical axis of the proper
distribution and the Gaussian approximation, which is shown in the left plot in 
Fig.~\ref{fig:e30}. We observe that the approximation is smaller near the center,
and is larger in the tails of the distribution. The latter is visible on the right 
plot, which shows the distribution on a logarithmic scale. Larger values in the
tails causes the rms to be biased to a larger value. We conclude that the following
heuristic equation may serve as a suitable approximation to estimate the luminosity
loss from filamentation caused by coupling errors
\begin{equation}\label{eq:lumireal}
{\cal L}/{\cal L}_0 = \frac{1}{\sqrt{1+0.34\kappa^2\eps_y/\eps_x}}
  \frac{1}{\sqrt{1+0.34\kappa^2\eps_x/\eps_y}}\ .
\end{equation}
We need to point out that we tested the approximation also for the other emittance
ratios in Fig.~\ref{fig:sumsum} with good success, at least up to $\kappa\approx 0.3.$
%..............................................
%..............................................
%
%%%%%%%%%%%%%%%%%%%%%%%%%%%%%%%%%%%%%%%%%%%%%%%%%%%%%%%%%%%%%%%%%%%%%%%
%
\section{CLIC}\label{sec:clic}
In Ref.~\cite{ANDREA} the effect of RF-breakdown in the accelerating structures of CLIC
on the beam is discussed and the quoted values for their magnitude are
\begin{itemize}
\item Transverse kick: $\theta\approx 29\,$keV
\item Focal length: $f\approx 5\,$m at 180\,MeV electron energy
\item Energy loss: $\Delta E\approx 23\,$MeV
\end{itemize}
where the energy loss corresponds to one non-contributing accelerating structure, because 
the breakdown created a plasma that reflects the incident RF.
\par
Part of the discussed analysis was used in ref.~\cite{AP2} to estimate the effect of 
breakdown on the CLIC beam. There it was found that the effect is most severe at 
the injection energy, where vertical kicks can have magnitude above 10 times the
angular divergence, leading to an annular beam that reduces the luminosity by as much
as a factor of 10 as is obvious from Fig.~\ref{fig:lumi}. The effect of energy loss
was found to be negligible.
\par
The focal length found to be 5\,m at 180\,MeV scales to about 250\,m at 9\,GeV
injection energy, which makes the factors describing the emittance increase rather 
small. Assuming beta-functions on the order of 10\,m we find $\beta^2/2f^2$ for 
quadrupolar errors and $\beta_x\beta_y/f^2$ for skew quadrupolar errors to be on 
the order of a few times $10^{-3}.$ This leads to negligible luminosity loss 
according to Fig.~\ref{fig:bblumi} and Fig.~\ref{fig:sumsum}.
%
%%%%%%%%%%%%%%%%%%%%%%%%%%%%%%%%%%%%%%%%%%%%%%%%%%%%%%%%%%%%%%%%%%%%%%%
%
\section{Conclusions}
We discussed the effect of perturbations to the beam due to RF-breakdown on the
geometric luminosity. We modeled the effect by assuming that the perturbations fully 
filament during the passage of the beam along the linear accelerator and calculated
the emittance increase as well as the shape of the resulting filamented distribution 
at the interaction point.
Based on the well-known results for transverse kicks and beta-function mismatch as
well as the, to our knowledge, new result for skew quadrupolar mismatch we used the
filamented distributions at the interaction point to convolute them with the 
beam traveling in the opposite direction and derived luminosity loss functions
for the different perturbations.
\par
For CLIC parameters we found that the most significant perturbations are vertical
kicks at the beginning of the linear accelerator where the beam energy is still
comparatively low. Quadrupolar errors from upright and skew-quadrupoles are less 
important.
\par
The application of the mismatch formulae is not limited to the determination of
luminosity loss. Injection errors into a ring will equally filament out and lead 
to an increased beam phase space, whose spatial profiles are given by those
shown in Fig.~\ref{fig:proj},~\ref{fig:bbdist} and~\ref{fig:distris}. Moreover,
static errors, such as stray fields incorrecty set power supplies for dipoles
and quadrupoles in the accelerator can be treated in the same way.
\par
% Discussions with A. Palaia who contributed to the evaluation of the CLIC lattice 
% discussed in section~\ref{sec:clic} are gratefully acknowledged. 
Support from the Swedish Research Council, contract 2014-6360 is acknowledged.
%
%
%%%%%%%%%%%%%%%%%%%%%%%%%%%%%%%%%%%%%%%%%%%%%%%%%%%%%%%%%%%%%%%%%%%%%%%
%
\bibliographystyle{plain}

\end{document}